\documentclass[preprint]{aastex}

\usepackage{graphics}
\newcommand{\bfig}{\noindent\begin{minipage}{3.48in}}
\newcommand{\efig}{\bigskip\end{minipage}}

\begin{document}

\title{The Variable Light Curve of GRB 030329: The Case for
Refreshed Shocks}



\author{Jonathan Granot}
\affil{Institute for Advanced Study, Olden Lane, Princeton, NJ
08540 ; granot@ias.edu}
\author{Ehud Nakar and Tsvi Piran}
\affil{Racah Institute for Physiscs, The Hebrew University,
Jerusalem 91904, Israel ; udini@phys.huji.ac.il,
tsvi@phys.huji.ac.il}

\begin{abstract}

GRB 030329 is unique in many aspects. It has a very low redshift
for a GRB, $z=0.1685$, and is therefore very bright and easy to
monitor, making it the most well studied afterglow to date. It
shows a supernova bump in the light curve, with a spectrum very
similar to SN 1998bw, thus establishing with much better
confidence the connection between GRBs and core collapse SNe.
There are also two important physical characteristics that make
this burst especially interesting, aside from its remarkably low
redshift. First, unlike most GRB afterglows, the light curve of
GRB 030329 shows a very large variability a few days after the
burst. These fluctuations show a roughly constact amplitude, and a
constant duration $\Delta t$, while $\Delta t/t$ decreases with
time $t$. Second, its $\gamma$-ray energy output and X-ray
luminosity at $10\;$hr are a factor of $\sim 20$ and $\sim 30$,
respectively, below the average value around which most GRBs are
narrowly clustered. We consider several interpretations for the
variability in the light curve, in the context of different
physical mechanisms, and find that the most likely cause is
refreshed shocks, i.e. slow shells that are ejected from the
source and catch up with the afterglow shock at late times. In GRB
030329 this happens after the jet break, which implies an
approximately constant duration $\Delta t$ of the bumps, in
agreement with the observations. This interpretation also
explains the anomalously low initial energy of this burst, as the
total energy of the afterglow shock is increased by a factor of
$\sim 10$ due to the refreshed shocks, thus bringing the total
energy output close to the average value for all GRBs.

\end{abstract}

\section{Introduction}
\label{introduction}

The afterglow light curve of most GRBs shows a smooth power law
decay with the time after the burst. This is consistent with the
prediction of the simplest model of a spherical blast wave
propagating into a uniform medium, where the spectrum consists of
several power law segments in which
$F_\nu\propto\nu^{-\beta}t^{-\alpha}$ (Sari, Piran \& Narayan
1998), and remains valid for an external density that drops as
$r^{-2}$ as expected for the stellar wind of a massive star
progenitor (Chevalier \& Li 2000) and for a jetted outflow (Rhoads
1999; Sari, Piran \& Halpern 1999).

However, a few GRBs do show deviations from a smooth power law
light curve. GRB 970508 showed a brightening at $t\sim 2\;$days,
which was observed in optical (Galama et al. 1998) and possibly
also in X-rays (Piro et al. 1998). An achromatic bump was observed
in the afterglow of GRB 000301C at $t\sim 4\;$ days (Berger et al.
2000; Sagar et al. 2000) and was attributed, among other
interpretations, to gravitational microlensing (Garnavich, Loeb \&
Stanek 2000). In these two cases a single epoch of re-brightening
was observed. In contrast to this, GRB 021004 displayed a highly
variable light curve, with several wiggles and deviations from a
simple power law behavior. This variable light curve was
interpreted as arising either due to a variable external density
(Lazzati et al. 2002; Nakar, Piran \& Granot 2003; Heyl \& Perna
2003) or due to a variable energy in the observed portion of the
afterglow shock. The latter can arise due to an irregular initial
distrubotion of the energy per unit solid angle in the outflow
(the patchy shell model, e.g. Kumar \& Piran 2000b). The early
slow decay of GRB 021004 was interpreted by Fox et al. (2003) as
arising from delayed shocks and continuous energy ejection from
the source (Rees \& M\'esz\'aros 1998; Kumar \& Piran 2000a; Sari
\& M\'esz\'aros 2000).

GRB 030329 was detected by HETE-II on March 29, 2003 (11:37:14.67
UT), and its location was distributed (Vanderspek et al., 2003) $t=1.4\;$hr
later. Optical observations from ground based telescopes soon
followed, and a bright optical afterglow with R=12.6 was detected
at $t=1.5\;$hr (Peterson \& Price 2003, Torii 2003). Observations at many
different wavelengths followed, and a very bright afterglow
emission was detected in the radio, sub-millimeter, millimeter and
X-ray (Marshall \& Swank 2003, Berger, Soderberg \& Frail 2003,
Hoge et al. 2003, Kuno, Sato \& Nakanishi 2003). Its redshift, $z=0.1685$
(Greiner et al. 2003),  places it as the closest GRB whose redshift was
firmly established\footnote{GRB 980425, if indeed associated with
SN 1998bw, was at a still lower redshift, of $z=0.0085$.}. It is
therefore not surprising that the afterglow of GRB 030329 is by
far the brightest afterglow detected so far\footnote{Here we
exclude the prompt optical emission which was observed for GRB
990123 (Akerlof et al. 1999) and peaked at tens of seconds after
at 9th magnitude, which is attributed to the reverse shock (Sari
\& Piran 1999; M\'esz\'aros \& Rees 1999).}, and is also the most
well studied GRB afterglow to date.

A spectral signature very similar to SN 1998bw began to emerge in
the spectrum of GRB 030329 after $\sim 7\;$days (Stanek et al.
2003), and became more prominent in the following days. This
provides the best evidence so far for the association of GRBs with
core collapse supernovae of the type of SN 1998bw, which are also
called hypernovae. The optical light curve showed a jet break at
$t=0.45\;$days, and several re-brightenings between $\sim 1.5$ and
$\sim 6\;$days, each with a duration of $\sim 0.3-0.8\;$days. We
explore different possible explanations for  the variability in
the light curve using different physical mechanisms and show that
the most likely explanation for the observed variability is
refreshed shocks.

\section{Interpretation of the Variable Light Curve}
\label{interp}

The optical light curve of GRB 030329 steepened from
$\alpha_1=0.93\pm 0.04$ to $\alpha_2=1.90\pm 0.03$ at $t_j\approx
0.45\;$ days (http://space.mit.edu/HETE/Bursts/GRB030329/). The
most likely explanation for this smooth change in the temporal
decay index $\alpha$  is a jet break, and in the following we
shall assume that this is indeed the case. Since the jet break is
quite sharp, a roughly constant (underlying) external density is
required. A stellar wind profile would not produce a sufficiently
sharp jet break (Kumar \& Panaitescu 2000).

The fluence GRB 030329 in the $15-5000\;$keV range was measured by
Konus-Wins to be $1.6\times 10^{-4}\;{\rm erg\; cm^{-2}}$ (Golenetskii
et al. 2003), which for its redshift of $z=0.1685$ and using the
WMAP cosmological parameters, implies an isotropic equivalent
$\gamma$-ray energy output of $E_{\gamma,{\rm iso}}\sim 1\times
10^{52}\;$erg. The X-ray flux in the $2-10\;$keV range was
measured by RXTE at $t=5\;$hr to be $F_X=1.4\times 10^{-10}\;{\rm
erg\; sec^{-1}\; cm^{-2}}$, which implies an isotropic equivalent
X-ray luminosity at $t=10\;$hr of $L_{X,{\rm iso}}\sim 5\times
10^{45}\;{\rm erg\;sec^{-1}}$ (Marshall \& Swank 2003). Using the same
formula and the  fiducial parameter values of Frail et al. (2001)
we find $\theta_{j,0}=0.07$ which corresponds to a beaming factor
of $f_b\approx\theta_{j,o}^2/2\approx 1/400$. This implies a true
$\gamma$-ray output of $E_\gamma\sim 3\times 10^{49}\;$erg, and a
true X-ray luminosity at $t=10\;$hr of $L_X\sim 1\times
10^{43}\;{\rm erg\; sec^{-1}}$. Both values are at the very low
end of the standard distribution for GRBs: Frail et al. (2001)
find $\log_{10}(E_\gamma[{\rm erg}])=50.7\pm 0.5$, while Berger,
Kulkarni \& Frail find $\log_{10}(L_X)=44.5\pm 0.5$.

The optical light curve shows  a well monitored re-brightening
between $t\sim 1.3$ and $\sim 1.7\;$days. This re-brightening
begins to effect the light curve after about one day, which is
only a factor of $\sim 2$ in time after $t_j$. Therefore, we
expect that the asymptotic decay slope $\alpha_2$ has not yet
been reached, and the value of $1.9$ inferred from observations
may be regarded as a lower limit. At least 3 additional and
similar features appear in the light curve at $t\sim
2.4-2.8\;$days, $t\sim 3.1-3.5\;$days and t$\sim 4.9-5.7\;$days,
though the latter are not as densely monitored (see Figure 1).
Another possible flattening of the light curve is observed at
$t\sim 9-10\;$days, however this may be at least in part due to
the contribution of an the underlying SN 2003dh, and is therefore
not included in our analysis. If the amplitude of each bump is
measured as the offset between the decay epochs before and after
the rise in the flux (which is roughly constant, as the temporal
decay indices $\alpha$ after the different bumps are rather
similar), we find that the amplitude of the first bump is a
factor of $\sim 2$, while the amplitude of the following bumps is
$\sim 1.3-1.4$. The exact determination of the amplitude depends
on the assumption for the underlying ``average" flux, and may
vary somewhat.

\subsection{Variable External Density}
\label{n_ext}

Variations in the external density was the first mechanism
proposed to account for variability in the light curve of GRB
afterglows. The variability due to small amplitude density
fluctuations caused by interstellar turbulence was considered by
Wang \& Loeb (2000). A density bump in the external medium was
suggested as a possible explanation for the bump in the light
curve of GRB 000301C (Berger et al. 2000), and a highly variable
external medium was suggested as an explanation for the large
variability in the afterglow of GRB 021004 (Lazzati et al. 2002;
Nakar, Piran \& Granot 2003; Heyl \& Perna 2003). A variable
external density produces a much larger variability below the
cooling break frequency $\nu_c$ (typically optical or IR),
compared to above $\nu_c$ (typically X-ray). For a spherically
symmetric external density distribution, it cannot produce sharp
variations in the flux on time scales $\Delta t<t$ (Nakar \& Piran
2003), due to angular spreading (i.e. the spreading in the arrival
time of photons to the observer due to the curvature of the
shock).

Turning to GRB 030329, we find that the rise in the normalization
of the decaying part of the light curve after each bump implies
that the external density would have to increase with radius, and
to do so in discrete steps, rather than in a smooth manner.
However, such a density profile seems very unlikely. If such a
density profile would still somehow occur, then the observed
optical light curve could be reproduced, if the jet would hardly
expand sideways at $t>t_j$. In this case, we expect much smaller
fluctuations in the X-ray light curve. As of yet, there is no good
X-ray light curve available that would enable us to compare the
variability in the light curve below and above $\nu_c$.

\subsection{Patchy Shell}
\label{patchy_shell}

The patchy shell model (Kumar \& Piran 2000b) suggests a
variability in the energy per unit solid angle of the GRB outflow,
on some typical angular scale. According to this model, the
amplitude of the fluctuations in the afterglow light curve is
expected to decrease with time $\propto\gamma$ (Nakar, Piran \&
Granot 2003). However, in GRB 030329 the bumps occur after the jet
break, when all the jet is visible, so that fluctuations in the
energy per unit solid angle across the jet should no longer
induce significant fluctuations in the light curve. Therefore,
assuming that the smooth steepening in the light curve at
$t=0.45\;$ days is indeed a jet break, the patchy shell model can
be ruled out as the source of the variability at $t>1\;$day.

\subsection{Refreshed Shocks}
\label{ref_shocks}

According to the refreshed shocks scenario, slow moving shells
that were ejected from the source with a modest Lorentz factor
catch up with the afterglow shock at late times well after the
internal shocks that are responsible for the prompt $\gamma$-ray
emission have ceased (Rees \& M\'esz\'aros 1998; Kumar \& Piran
2000a; Sari \& M\'esz\'aros 2000). Each such shell injects energy
into the afterglow shock as it collides with it from behind, and
causes a re-brightening in the afterglow light curve. After the
re-brightening the same power law decay as before the bump should
be resumed, as long as the observed frequency remains within the
same power law segment of the spectrum (i.e. if no break frequency
passes through the observed band). Similarly, under these
conditions the color indices should not change. The normalization
of the decaying light curve after the bump is higher than before
the bump by a factor $f$, due to the increase in the energy of
the afterglow shock. The energy increases by a factor  $f^{4/(3+p)}$
for $\nu_m<\nu<\nu_c$ and
$f^{4/(2+p)}$ for $\nu_m,\nu_c<\nu$. For GRB 030329 $p\approx 2$
and $\nu_c(t_j)$ is around the optical, so that the increase in
the energy is approximately linear in $f$. We find that
across all the bumps in the first several days, this normalization
increased by a factor of $f_{\rm total}\sim 10$, and deduce that
this implies a similar increase in the energy of the afterglow
shocks due to energy injection from the shells that caught up with
it. The conservative estimate of $f_{\rm total}$ using the value
of $\alpha_2$ from just after $t_j$ is $\sim 5$, while if we take
$\alpha_2$ from the decay after the second or third bumps we
obtain $f_{\rm total}\sim 20$ (see Figure 1). Therefore $f_{\rm total}\sim 10$
with an uncertainty of a factor $\sim 2$ seems like a reasonable
estimate.

In the original refreshed shocks scenario it was expected that
these shocks occur before the jet spreading, and therefore $\Delta
t\sim t$ (Kumar \& Piran 2000b). However, in GRB 030329 the jet
break is at $0.45\;$days, and the refreshed shocks took place at a
later time, after the whole jet becomes visible as its Lorentz
factor is smaller than the inverse of its opening angle. We expect
that all the shells ejected from the central source will have
roughly the same opening angle, $\theta_{j,0}$ (see illustration
in Figure 2). The most forward shell sweeps up the ambient medium
and produces the afterglow emission. Since it is very hot, with an
internal energy much higher than its rest energy, it could
potentially expand sideways at close to the speed of light $c$ (or
$c/\sqrt{3}$) in its local rest frame, though the lateral
expansion could still be much smaller than this. However, the
slower shells that follow in the wake of the forward shell
encounter very little material before they catch up with the
forward shell, and  therefore are expected to be cold, and should
not expand sideways significantly before colliding with the
afterglow shock. In this case the duration of the bump in the
light curve is given by the angular time $\Delta t\sim
R\theta_{j,0}^2/2=t_j(R/R_j)=t_j(t/t_j)^a$, where $a=1/4$ if the
lateral spreading of the jet is negligible, while $a\approx 0$ if
the lateral spreading is at the local sound speed. This allows
$\Delta t <t$. Numerical simulations suggest that the lateral
spreading is quite modest (Granot et al. 2001), implying $a\approx
1/4$. This is consistent with the almost constant duration of the
first 3 bumps, and the slightly larger duration of the fourth
bump, that are seen in GRB 030329.

The temporal decay slope after the first bump is very close to
$\alpha_2$, while after the second and third bumps $\alpha$ is
slightly larger than $\alpha_2$. Unfortunately, it is hard to
accurately determine the value of $\alpha$ after the forth bump
due to the contribution from the underlying SN, though it seems
similar to the values for the preceding bumps. The steeper
temporal decay slopes, $\alpha>\alpha_2$, may be due to the fact
that the asymptotic power law decay after the jet break was not
reached, and is underestimated by $\alpha_2$. The small
differences between the slopes may be attributed to a slight
inhomogeneous distribution of the Lorentz factor across the slow
shell that catches up (i.e. its Lorentz factor decreases smoothly
from its front edge to its back edge, so that the energy
injection is not step-like but more gradual). We conclude that
the refreshed shocks model provides the best explanation for the
variability in the light curve of GRB 030329. The timing of the
bumps suggest that the Lorentz factors of the slower shells
ranged from $\ge 6$ to $\ge 3.5$.  The lower limit is obtained
for the extreme case of when after the jet break the jet expands
sideways at the  the local speed of sound  and its radius  is
almost constant (Rhoads 1997). The higher values are obtained if
the jet does not spread sideways significantly, as suggested by
the results of Granot et al. (2001). In this case the radius of
the front shell increases with time and the inner shells need a
larger Lorentz factor to catch up with the forward one.

\section{Discussion}
\label{discussion}

We have compared the variable light curve of GRB 030329 to the
predictions of the different models for variability in GRB
afterglows and find that the only satisfactory explanation for the
observed variability in GRB 030329 is refreshed shocks that take
place after the jet break. This interpretation is supported by
the almost constant durations of the different bumps, $\Delta t
\sim 0.4-0.8\;$days, which is also of the order of the jet break
time, $t_j=0.45\;$days, and by the step-wise shape of the light
curve. According to the refreshed shocks interpretation, we find
that the energy injected into the afterglow shock by all the slow
shells that catch up with it, increases its initial energy by a
factor of $\sim 10$. This implies that the total energy in this
GRB outflow is an order of magnitude larger than the energy
inferred from the prompt $\gamma$-ray emission, or from the early
afterglow emission before $\sim 1\;$day, and thus explains the
anomolously low values inferred for $E_\gamma$ and $L_X$. A
direct prediction of this interpretation is that we should expect
very significant  radio flares  corresponding to the observed
optical bumps. These should arise from emission by the reverse
shocks that form in the refreshed shocks.

\acknowledgments

This research was supported in part by funds for natural sciences
at the Institute for Advanced Study (JG), by a grant from  the
Israel Space Agency (EN and TP) and by a grant from the Horowitz
foundation (EN).


\begin{figure}[htb]
\begin{center}
\resizebox*{1.\columnwidth}{0.5\textheight}{\includegraphics{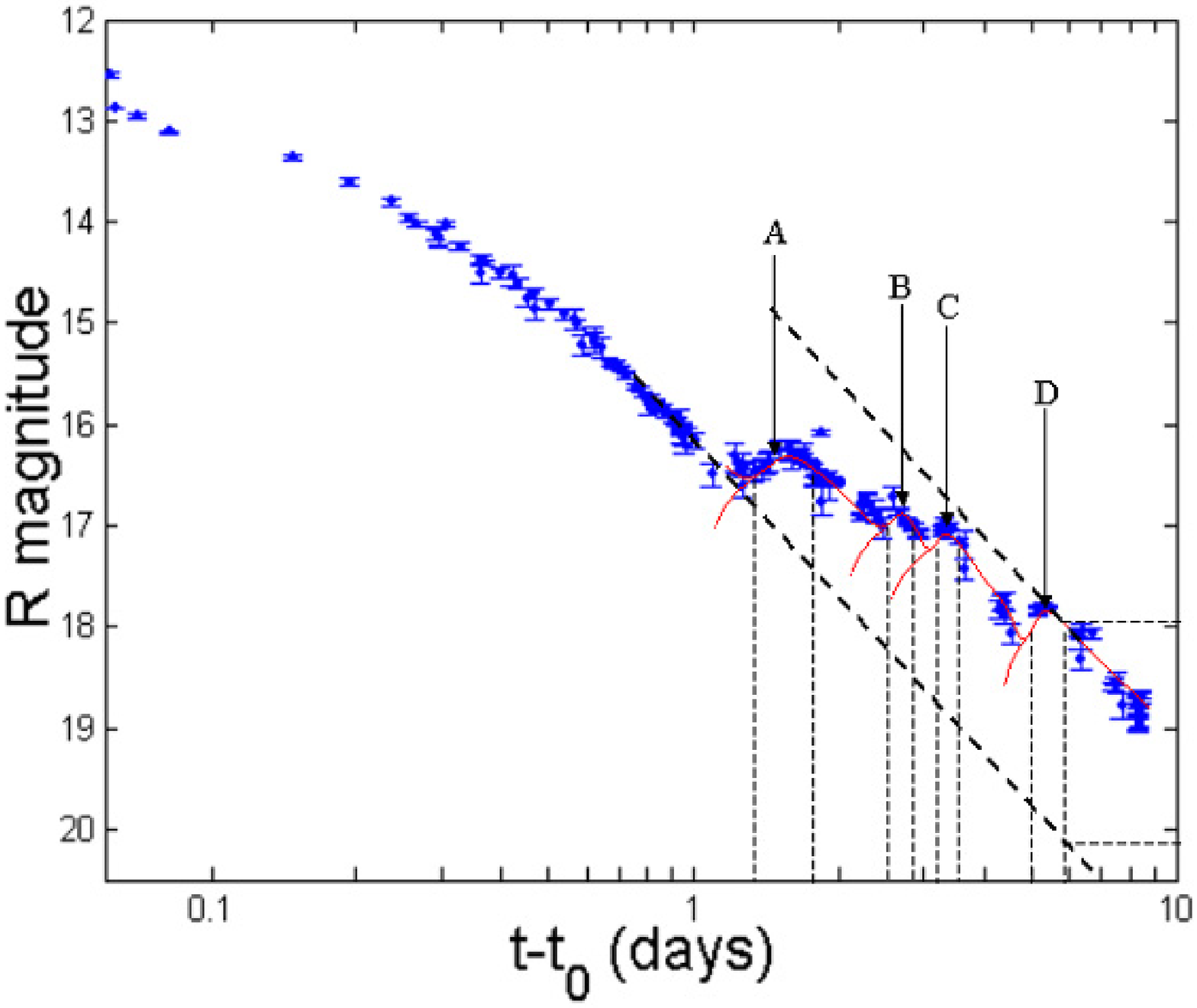}}
\caption{The GCN light curve of GRB 030329 during the first 9
days. The vertical dashed lines mark the position of the four
bumps (A at  $t\sim 1.3-\sim 1.7\;$days, B at  $t\sim
2.4-2.8\;$days, C at $t\sim 3.1-3.5\;$days, and D at $t\sim
4.9-5.7\;$days). The thick slanted lines mark the asymptotic decay
slopes before and after the bumps. The horizontal dashed lines
mark the difference in magnitude between these two curves. The
thin lines depicts the schematic shapes of an individual refreshed
shock light curve. The data points are taken from the following
GCNs: Rykoff et al. 2003; Burenin et al. 2003a,b,c,d,e; Pavlenko
et al. 2003a,b,c; Bartolini et al. 2003; Price, A. et al.
2003a,b,c; Weidong et al. 2003; Martini et al. 2003; Zharikov et
al. 2003a,b; Brodney et al. 2003a,b; Halpern  et al. 2003; Lipkin
et al. 2003a,b,c; Rumyantsev et al. 2003; Klose et al. 2003;
Stanek et al. 2003 ; Zeh et al. 2003; Ibrahimov et al. 2003a,b,c;
Cantiello et al. 2003; Sato et al. 2003; Khamitov et al.
2003a,b,c. An arbitrary errorbar of 0.1mag is taken for any data
point with no error included in the GCN. The collection of data
from the GCN at the the first 80 hours is taken from
http://astron.berkeley.edu/~bait/grb/gcn030329.r.dat. }
\end{center}
\end{figure}

\begin{figure}[htb]
\begin{center}
\resizebox*{1.\columnwidth}{0.7\textheight}{\includegraphics{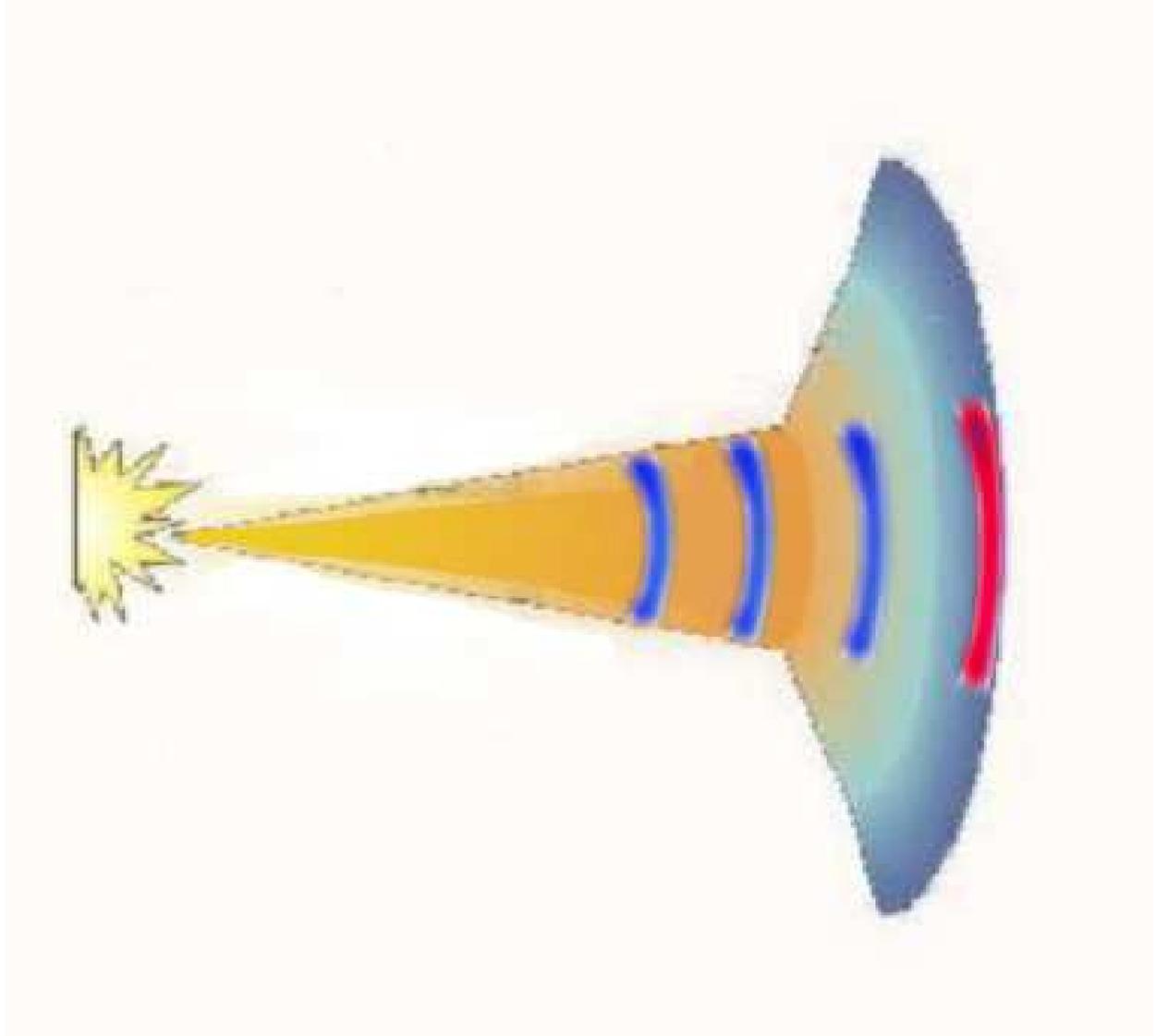}}
\caption{A schematic diagram that describes the refreshed shocks
scenario after a jet break. The original flow is collimated to an
opening angle $\theta_{j,o}$. At $t \sim R \theta_{j,o}^2/2c$ the
Lorentz factor reaches $\gamma \sim \theta_{j,o}^{-1}$ and begins
spreading sideways. The following shells, which propagate in the
wake of the forward shell, remain cold and do not spread sideways
until they collide with the forward shell and a refreshed shock
forms. The duration of the brightening (which is dominated by the
angular spreading time) is determined by the angular size of the
slow shells, $\sim R\theta_{j,o}^2/2c$, which is comparable to
the jet break time.
 }
\end{center}
\end{figure}

\end{document}